\documentclass[conference]{IEEEtran}

\usepackage{fancyhdr}
\usepackage{cite}
\usepackage{amsmath,amssymb,amsfonts}
\usepackage{algorithmic}
\usepackage{graphicx}
\usepackage{textcomp}
\usepackage{xcolor}
\def\BibTeX{{\rm B\kern-.05em{\sc i\kern-.025em b}\kern-.08em
    T\kern-.1667em\lower.7ex\hbox{E}\kern-.125emX}}

\usepackage{xspace}
 \usepackage{amsmath,amssymb,amsfonts}
 \usepackage{amsmath}
\usepackage[switch]{lineno}
\usepackage{algorithm} 
 \usepackage{algorithmic,tabularx}
 \usepackage{multirow}
 \usepackage{multicol} 
 \usepackage{array}
 \usepackage{booktabs}
 \usepackage{CJK}
 \usepackage{times}
 \usepackage{pifont}
 \usepackage{graphicx}
 \usepackage{textcomp}
 \usepackage{subfigure} 
 \usepackage{comment}
 \usepackage{adjustbox}
 \usepackage{ulem}

\newlength\dwqindent
\newlength\wqindent
\usepackage{anyfontsize}

\usepackage{ulem,xpatch}

\usepackage{color}
\usepackage{xcolor}

\usepackage{hyperref}
\usepackage{colortbl}
\usepackage{comment}

\definecolor{wq}{RGB}{23,125,54}
\definecolor{dong}{RGB}{125,125,0}
\definecolor{gokcen}{RGB}{255,50,255}
\definecolor{zhen}{RGB}{23,125,54}
\definecolor{revision}{RGB}{0,0,0}

\newcommand{\name}{Smart-PGSim\xspace}

\begin{document}





\title{\name: Using Neural Network to Accelerate AC-OPF Power Grid Simulation}

\author{\IEEEauthorblockN{Wenqian Dong}
\IEEEauthorblockA{\textit{UC Merced} \\
California, USA\\
wdong5@ucmerced.edu }
\and
\IEEEauthorblockN{Zhen Xie}
\IEEEauthorblockA{\textit{UC Merced}\\
California, USA \\
zxie10@ucmerced.edu}
\and
\IEEEauthorblockN{Gokcen Kestor}
\IEEEauthorblockA{\textit{Pacific Northwest National Laboratory} \\
Washington, USA \\
gokcen.kestor@pnnl.gov}
\and
\IEEEauthorblockN{Dong Li}
\IEEEauthorblockA{\textit{UC Merced}\\
California, USA\\
dli35@ucmerced.edu}
}

\maketitle

\thispagestyle{fancy}
\lhead{}
\rhead{}
\chead{}
\lfoot{\footnotesize{
SC20, November 15-20, 2020, Atlanta, GA, USA
\newline 978-1-5386-8384-2/XX/\$31.00 \copyright 2020 IEEE}}
\rfoot{}
\cfoot{}
\renewcommand{\headrulewidth}{0pt}
\renewcommand{\footrulewidth}{0pt}

\begin{abstract}

In this work we address the problem of accelerating complex power-grid simulation through machine learning (ML). Specifically, we develop a framework, \name, which generates multitask-learning (MTL) neural network (NN) models to predict the initial values of variables critical to the problem convergence. 
MTL models allow information sharing when predicting multiple dependent variables while including customized layers to predict individual variables.
We show that, to achieve the required accuracy, it is paramount to embed domain-specific constraints derived from the specific power-grid components in the MTL model.  \name then employs the predicted initial values as a high-quality initial condition for the power-grid numerical solver (warm start), resulting in both higher performance compared to state-of-the-art solutions while maintaining the required accuracy. \name brings $2.60\times$ speedup on average (up to $3.28\times$) computed over 10,000 problems, without losing solution optimality.
\end{abstract}


\section{Introduction}
Artificial Intelligence (AI) and Machine Learning (ML) techniques are revolutionizing the way researchers approach scientific and engineering problems. By employing reverse-engineering and automatic learning methodologies it is often possible to solve complex, unstructured problems with a fraction of the computing power and execution time required by traditional direct and first-principle methods. ML provides researchers with a powerful tool to learn the structure of physical phenomenon directly from Nature, rather than having to explain the causal relationships through direct application of physics law. Many research and engineering fields, from image recognition to autonomous driving, from health to natural language processing (NLP), have experienced a tremendous boost in performance and efficiency over the last few years. Many problems that seemed impossible to be solved, can now be tackled thanks to the use of ML methodologies.

The use of ML methodologies in scientific and engineering applications has been, somehow, limited. By using Neural Network (NN) as a tool to learn and model complicated (non-)linear relationships between input and output data sets, scientists have shown preliminary success in some HPC problems (e.g., detecting neutrinos~\cite{ichep16:radovic}, climate simulations~\cite{nips17:racha}, and fluid dynamic simulation~\cite{dong2019adaptive}). 
With NN, scientists are able to augment existing scientific simulations by improving simulation accuracy and significantly reducing latency~\cite{richard2014artificial, sr15:widley, mathuriya2018cosmoflow, balaprakash2019scalable, yang2019highly,liu2019tomogan,liu2019deep}. However, although there have been successful studies of applied ML to scientific applications, these fields have not experienced the double- or triple-digit improvements seen in other domains. The reason for such discrepancy is the fundamentally different characteristic of scientific and engineering applications compared to domains such as image recognition and NLP: scientific applications require a level of precision and robustness that may not be provided by most of the current ML methods employed in other domains.

In this work we study the implication of using ML techniques to accelerate the power-grid simulations, the structure of the ML model to be used, the relative importance of the features selected, and, most importantly, the impact of incorporating physics constraints on the performance of the application.
The power-grid simulation~\cite{liu2010optimal} is a complex nonlinear optimization problem for the management of power flow and is critical to the power industry in electricity dispatch scheduling, reliability analysis, and maintenance planning for power and generators~\cite{fernandez2013real, chen2001efficient}.  
The alternating current optimal power flow (AC-OPF) simulation is the most fundamental and time-consuming part of the power grid simulation. The problem size of AC-OPF is generally large, in which the scale of the generator node can vary from $10^3$ to $10^6$~\cite{huang2002parallel, guo2000nonlinear, momoh1999improved}. 
Despite the large problem size, the AC-OPF simulation requires near real-time updates during power scheduling.  
In a typical scenario, power grid operators repeatedly solve the optimal power flow problem multiple times within every minute throughout a day, every day of the year~\cite{savulescu2014real, petra2014real,din2017short}, for decades, to ensure that the power grid system is operating reliably and safely. The high requirement on the simulation latency and frequent usage of the simulation make the power grid simulation a mission-critical application \textcolor{revision}{under active development in the HPC community and within the U.S. Department of Energy (DOE) and the DOE Exascale Computing Project (ECP)}.

NN has been applied to solve the optimal power flow problem in the past~\cite{guha2019machine, baker2019learning, zamzam2019learning, deka2019learning, ng2018statistical}. However, existing efforts have focused on improving performance by entirely replacing the simulation solver with an approximated NN model or facilitating existing solvers by identifying active constraints.
While these approaches provide considerable speedups, NN provides only an \textit{approximation} of the optimal solution or approximates computation in the simulation. As a result, these approaches may not provide the desired precision for the solution or may provide a non-optimal solution. In the context of power-grid simulation, the first case results in an infeasible solution (e.g., not being able to provide the required power to satisfy the user demand) while the second case may results in a large economic loss (i.e., solving the problem at a much higher cost.)

In this paper, we introduce a new method to apply NN to the AC-OPF simulation. 
Unlike the existing studies, we employ NN to generate an initial solution and then inject it to the AC-OPF solver. 
Because of the high quality of the initial solution and guidance of other outputs generated by the proposed NN, the simulation can run faster (or converge faster) without losing the solution optimality. 

There are several challenges to apply our method to the AC-OPF simulation. 
First, deciding which variables in the AC-OPF simulation should be used as NN output and quantifying the sensitivity of simulation execution time and convergence to those variables is a challenge. The AC-OPF simulation involves a set of variables, including power grid information and multiple variables critical for the computation convergence. We cannot use all of them as NN output because that largely increases network complexity and puts high requirements on training efficiency and sufficiency of training samples. On the other hand, using only the solution of the AC-OPF as NN output, we often lose simulation robustness because of the limited guidance for the simulation from the initial solution. Furthermore, understanding the sensitivity of simulation time and convergence to those variables is useful for deciding NN topology and generating high-quality initial solutions.

Second, how to apply NN to the AC-OPF simulation without disturbing the simulation robustness is a challenge. 
Due to the non-convex and nonlinear nature of the AC-OPF problem, the simulation process itself is at the risk of a failed convergence with the use of iterative numerical methods. 
The simulation must be robust enough to handle various power flow cases with computation convergence. Using NN to generate an initial solution, we must make sure that the initial solution makes sense and does not impact the computation convergence in the original simulation.

Third, how to impose physical constraints on NN to ensure the validness of NN prediction. 
Traditionally, the NN model is manually constructed by computer scientists as a black box with limited or no domain knowledge and without considering domain requirements. Although NN models can be adjusted as a nonlinear tool box to accommodate a change of inputs and generate some approximation, the understanding of the model is lost. 
Instead of blindly trusting that the data mining algorithm will produce a correct model, we seek for what variables physically mean and which physical laws are driving the interpretable evolution of the analysis paradigm.

To address the above challenges, we introduce, \name, a framework that facilitates the construction of a NN model to accelerate the AC-OPF simulation. \name is based on the following design principles. First, it generates an NN model that uses power grid components as \textcolor{revision}{inputs} and variables critical for the simulation convergence as the model output.
By using \name, we perform a sensitivity study to understand the impact of the output accuracy on execution time and convergence, by using precise or imprecise data for some variables. 
This sensitivity study provides guidance  on choosing a correct and efficient NN topology.

Second, \name uses a novel multitask-learning NN model to accelerate the AC-OPF simulation. The model topology allows information sharing when predicting multiple dependent variables while including customized layers \textcolor{revision}{for each} variable. This multi-task model improves the model accuracy, compared with the traditional single-task model, while simplifying the training process.

Third, \name allows embedding physical constraints from the original formulation \textcolor{revision}{of} the AC-OPF problem into the NN model and imposes those constraints into the training objective function or the last layer based on transformation on equality and inequality in the constraints. 

We summarize our major constitutions as follows.
\begin{itemize}
\item A systematic approach and a framework (\name) to accelerate optimization problems in general and the AC-OPF power grid simulation in particular; 
 
\item A set of techniques to construct NN models for robust, accurate, and high-performance numerical solvers; 

\item We show that \name achieves $2.60\times$ speedup on average (with the consideration of NN cost) and up to $3.28\times$ over the original AC-OPF simulation method (computed over 10,000 problems as the simulation input), without losing the optimality of the final solution.

\end{itemize}

 \section{Background}
\label{sec:acopf}

In this section, we review the problem formulation and the primal-dual interior-point method in the AC-OPF problem.

\subsection{Problem Formulation for AC-Optimal Power Flow}
The AC-OPF problem aims at minimizing an objective function by optimizing the power dispatch and transmission decisions. 
The objective function calculates the cost of power generation, subjecting to physical, operational, and technical constraints including  Kirchhoff’s laws, operating limits of generators, voltage levels, and loading limits of transmission lines~\cite{Johnson1997}.  
The standard AC-OPF problem is formulated as: 

\vspace{-11pt}
\begin{subequations} 
\begin{align}
 &\min_{X}f(X) \label{eq:1a}\\
s.t.\ &G(X)=0 \label{eq:1b}\\
 &H(X)>0 \label{eq:1c}   \\
 &X_{min} \le X \le X_{max}. \label{eq:1d} 
\end{align}
\end{subequations} 
where $f(X)$ is the cost function to be minimized, and $X$ is an optimization vector as the simulation solution. 
Eqn.~\ref{eq:1b} builds an equality constraint, which sets up power balance incorporating variable bounds. The formulation~\ref{eq:1c} is an inequality constraint that sets up branch flow limits. The optimization vector $X$ is bounded by $X_{min}$ and $X_{max}$ which introduces the constrains on reference bus angles, voltage magnitudes, and generator injections.    
The optimization vector $X$ consists of four variables, $X =\{ V_a; V_m; P_g; Q_g \}$, i.e., voltage angles $V_a$, voltage magnitudes $V_m$, generator real power injections $P_g$ and reactive power injections $Q_g$.

In power grid simulation, $G(X)=0$ is an AC nodal power balance equation and enables the AC-steady conditions of the power system, which can be split into real and reactive parts:

\vspace{-11pt}
\begin{subequations}
\label{eqn:3}
\begin{align}
    P_i(C_g,P_g) = P_d + P_{bus}(Y_{bus}, V_a, V_m)\\
    Q_i(C_g,Q_g) = Q_d + Q_{bus}(Y_{bus}, V_a, V_m).
\end{align}
\end{subequations}
In Eqn.~\ref{eqn:3}, $C_g$ is the generator connection matrix reflecting generator locations in a power grid network. $Y_{bus}$ is the bus admittance matrix including all constant impedance elements. $P_i$ and $Q_i$ refer to power real and reactive injection for the power system. $P_d$ and $Q_d$ are power loads. $P_{bus}$ and $Q_{bus}$ are power consumption of transmission lines.

\subsection{Primal-dual Interior Point Solver}
\label{interior_point_solver}
The primal-dual interior point method~\cite{mehrotra1992implementation, wang2007computation} is an efficient algorithm to solve the non-convex optimization problem for AC-OPF. 
Matpower~\cite{zimmerman2016matpower} is a widely used framework for solving power flow and optimal power flow problems. Matpower uses a solver, called MIPS, to solve those problems.

To solve the AC-OPF problem, MIPS first converts the inequality constraint in Eqn.~\ref{eq:1c} into an equality constraint with a vector $Z$,  $H(X)+Z=0$ where $Z$ is a vector of positive slack variables.   
MIPS further uses a barrier function $\ln(Z)$ to bound $Z$. Based on that, MIPS uses a Lagrangian formulation to formulate the AC-OPF problem as follows.
\vspace{-8pt}
\begin{equation}
\label{eqn:largran}
\footnotesize
 L^{\gamma}(X,Z,\lambda,\mu)=f(X)+\mathbf{\lambda}^\intercal G(X)+\mathbf{\mu}^\intercal(H(X)+Z)-\gamma\sum_{m=1}^{n_i}\ln(Z_m)
\end{equation}

where $\lambda$ is called the equality Lagrangian multiplier, $\mu$ is called the inequality Lagrangian multiplier, and $\gamma$ is called the perturbation parameter. During the solving process, $\gamma$ is approaching zero.  If $\gamma=0$, the solution to this Lagrangian formulation equals to that of the original form (Eqn. 1).  

Matpower uses Newton method to solve Eqn.~\ref{eqn:largran}, which iteratively converges to a set of convergence criteria  (particularly four terminate conditions)~\cite{zimmerman2016matpower}. 
The Newton Method is computationally intensive and requires constant updates of input and output variables:  the method firstly updates $X$ and $\lambda$, then $Z$ based on $ X$, and, finally, $\mu$ based on $X$ and $ Z$. As we will see in the next Sections, this structure introduces internal dependencies on the variables that our model exploits for better performance and accuracy.

\section{Related Work}

OPF problems can be categorized into three forms: economic dispatch (ED)~\cite{cohen1997system}, Direct Current (DC-OPF)~\cite{Stott2009}, and Alternating Current (AC-OPF) problems~\cite{Christie2000}. 
The AC-OPF problem is the original OPF problem, which is non-convex and the most challenging one among the three. 
ED and DC-OPF problems are the relaxed version of the AC-OPF problem, which is obtained by removing or linearizing some constraints in the AC-OPF problem, respectively. 
Traditionally, numerical iteration algorithms are used to solve the OPF problem~\cite{SousaInterior2011,Low2014ConvexRO,Jabr2006Conic,Petra2014PIPS,lubin2013parallel}.
However, the time complexity of these algorithms might be significant, especially when the scale of the transmission power system becomes large.
To deal with this limitation, researchers have explored learning-based approaches to accelerate solving OPF problems.

Vaccaro et al.~\cite{Vaccaro2016pca} use the principal component analysis (PCA) to identify unknown relationships among OPF variables, which reduces the number of variables to be solved for a solution. Ng et al.~\cite{ng2018statistical} use a statistical learning-based approach to set up a mapping between input power requirement and output dispatch scheme. 
However, the approaches mentioned above consider only the prediction accuracy without taking into account the correlation among OPF problem variables, which leads to a solution that can not satisfy all of the problem constraints. Pan et al.~\cite{pan2019deepopf} use the multilayer perceptron (MLP) to learn the mapping between input and decisions for DC-OPF and apply it to obtain optimized operating decisions upon arbitrary inputs.
While this approach is effective for DC-OPF, it has low generalization capacity and cannot be applied to a non-convex problem such as AC-OPF.
Previous works~\cite{guha2019machine, baker2019learning, zamzam2019learning} have leveraged machine learning to accelerate the AC problem.
Zamzam et al.~\cite{zamzam2019learning} develop an online method based on machine learning to obtain feasible solutions to the AC problem by loading the optimal generator set-points and enforcing generation limits. However, the AC grid contains more voltage phase angles beyond magnitudes and reactive parts of power generation. Unlike these methods, the proposed approach includes all the inputs of the AC problem and guarantees that the predicted solution is optimal while providing significant performance improvement.

\begin{figure*}[!t]
  \centering
  \includegraphics[width=1.0\linewidth]{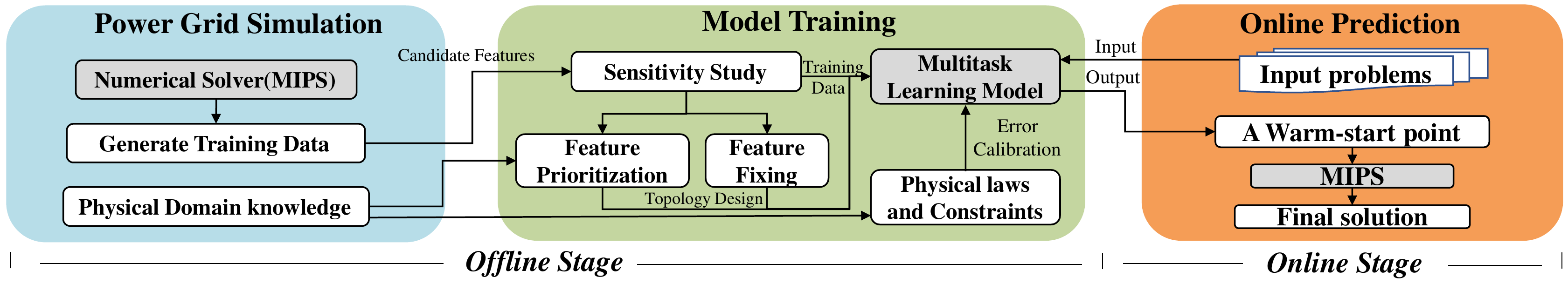}
  \caption{Workflow of the proposed Smart-PGsim}
  \label{fig:overview} 
\end{figure*}

In this work, we use NN models to solve the AC-OPF problem.
We follow a radically different approach compared to previous work in that we employ ML to estimate a high-quality initial solution for the solver, greatly speeding up the entire computation, and then leverage traditional AC-OPF solver to guarantee precision and robustness of the solution. We show in the next Sections that our approach can simultaneously provide large performance improvement \textit{and} high-precision solutions.

\section{Overview}
This section overviews our proposed framework ``Smart-PGsim".
The Smart-PGsim framework includes two phases: offline and online phases. 
Figure~\ref{fig:overview} shows the workflow of \name. 

The offline phase investigates the power grid simulation to find the most crucial features to construct an efficient NN model for online acceleration.
In particular, our sensitivity study (Section~\ref{sec:sensi_study}) \textcolor{revision}{firstly} identifies the most important features (in other words, determining variables in MIPS as the output of the prediction model) and quantifies the impact of the imprecise variables (i.e., variables with some accuracy loss) on the success rate of simulation and performance in terms of execution time. The results in sensitivity study are used to guide the NN topology design.

Then, \name constructs a multi-task learning (MTL) model (Section~\ref{sec:multitask}) guided by the sensitivity study.
The model shares domain information between prediction tasks, while uses a customized topology design for each task. \name prioritizes features to distinguish main tasks and auxiliary tasks and applies a physics-dependent hierarchy for those features have domain specific dependency.

Next, \name incorporates physical domain knowledge during model training to improve prediction quality (Section~\ref{sec:physics_info}).
The domain knowledge presents physical constraints providing explicit and implicit error bounds. 
Using the domain knowledge improves prediction accuracy, interpretability, and defensibility of the MTL model, while simultaneously augmenting physical data as complementary.

After the above offline phase, the \textcolor{revision}{well-trained} MTL model can be used to \textcolor{revision}{generate} a warm-start point for MIPS as online prediction. 
The MIPS (or other numerical solvers) can use these high-quality start points for quick convergence.

\section{Sensitivity Study}
\label{sec:sensi_study}

\begin{table*}
\small
\caption{\textbf{Ablation study on the input signals}  \protect\linebreak
\textmd{The SR and SU represent Success Rate(\%) and Speedup($\times$) respectively.}}

\label{tab:Ablation}
\begin{adjustbox}{width=1.0\textwidth,center}
\begin{tabular}{c cccc cc cc cc cc cc cc cc cc c}
       \toprule
\multirow{2}{*}[-0.33em]{} &
\multirow{2}{*}[-0.33em]{$X$} &
\multirow{2}{*}[-0.33em]{$\lambda$} &
\multirow{2}{*}[-0.33em]{$\mu$} &
\multirow{2}{*}[-0.33em]{$Z$} &
\multicolumn{2}{c}{bus 5}&  
\multicolumn{2}{c}{bus 9} & 
\multicolumn{2}{c}{bus 14} & 
\multicolumn{2}{c}{bus 30} & 
\multicolumn{2}{c}{bus 39} & 
\multicolumn{2}{c}{bus 57} & 
\multicolumn{2}{c}{bus 118} & 
\multicolumn{2}{c}{bus 300} &
\multirow{2}{*}[-0.33em]{Observation} \\

\cmidrule(lr){6-7}
\cmidrule(lr){8-9}
\cmidrule(lr){10-11}
\cmidrule(lr){12-13}
\cmidrule(lr){14-15}
\cmidrule(lr){16-17}
\cmidrule(lr){18-19}
\cmidrule(lr){20-21}
 &	&&&  &SR& SU  &SR& SU  &SR& SU  &SR& SU  &SR& SU  &SR& SU  &SR& SU  &SR& SU & \\
  		\midrule
\uppercase\expandafter{\romannumeral1} &0&0&0&0 &100&1&100&1&100&1&100&1&100&1&100&1&100&1&100&1&baseline\\ 
\uppercase\expandafter{\romannumeral2} &0&0&0&1 &0&--&10&0.67&0&--&89&0.66&0&--&6&0.99&0&--&0&-- &\\ 
\uppercase\expandafter{\romannumeral3} &0&0&1&0 &100&1.04&100&0.73&100&0.92&100&0.93&95&0.95&88&0.60&99&1.03& 100 &1.24 \\
\uppercase\expandafter{\romannumeral4} &0&0&1&1 &100&1.09&100&0.96&100&0.85&3&0.22&75&1.02&99&0.72&0&--& 68& 1.24 &\\
\uppercase\expandafter{\romannumeral5} &0&1&0&0 &100&0.98&100&0.99&100&1.00&30&1.06&100&1.00&100&0.98&100&0.98& 98 &  0.99&OBS 3\\
\uppercase\expandafter{\romannumeral6} &0&1&0&1 &0&--&8&0.73&0&--&79&0.70&0&0.61&9&0.90&0&--& 0&-- &\\
\uppercase\expandafter{\romannumeral7} &0&1&1&0 &100&0.99&0&--&80&0.61&98&0.90&100&1.09&100&0.89&100&0.93 & 100 & 1.08 &\\
\uppercase\expandafter{\romannumeral8} &0&1&1&1 &100&1.05&100&1.24&100&1.32&30&0.19&100&1.39&100&1.25&100&1.59 & 100 & 1.75& OBS 3\\
\uppercase\expandafter{\romannumeral9} &1&0&0&0 &100&1.17&100&0.99&100&0.99&100&0.95&100&1.06&100&1.01&100&0.99 &100 &1.08& OBS 1, 4\\
\uppercase\expandafter{\romannumeral10} &1&0&0&1 &0&--&0&--&0&--&0&--&0&--&11&0.94&0&-- & 0 &--& OBS 2, 4\\
\uppercase\expandafter{\romannumeral11} &1&0&1&0 &100&1.28&100&0.85&100&0.88&100&1.33&95&1.48&100&0.67&90&0.84& 96 &1.24 &OBS 4\\
\uppercase\expandafter{\romannumeral12} &1&0&1&1&100&1.45&100&1.03&100&0.80&95&1.26&80&1.22&100&0.65&0&--& 53&0.87 &OBS 2 \\
\uppercase\expandafter{\romannumeral13} &1&1&0&0&100&1.18&100&1.00&100&0.98&100&0.94&100&1.06&100&1.00&100&0.99 & 100 & 1.07&  OBS 3, 4\\
\uppercase\expandafter{\romannumeral14} &1&1&0&1&0&--&0&--&0&--&0&--&0&--&10&0.97&0&-- &100&1.18 & \\
\uppercase\expandafter{\romannumeral15} &1&1&1&0&100&1.28&100&0.79&100&0.90&100&1.09&100&1.22&100&0.93&100&0.94 & 100 &  1.32 &\\
\uppercase\expandafter{\romannumeral16} &1&1&1&1&100&5.21&100&4.58&100&3.74&100&6.15&100&6.60&100&4.58&100&7.63& 100 &  14.6 &OBS 1, 3\\
\bottomrule
\end{tabular}
\end{adjustbox}
\end{table*}



The ability of NN to produce high-quality results is the key to improve simulation performance (making the simulation quickly converged). In this section, we discuss the opportunity available in NN with the assist of a sensitivity study tool that detects and analyzes those variables critical to simulation convergence and execution time.


We introduce two data types, i.e., \textit{imprecise default data} and \textit{precise simulation data}, to study the impact of noisy feature to simulation quality and execution time. 
By doing so, we can check the (lowest) highest performance brought by these (im-)precise data, which demystifies the contribution of each feature to success rate and speedup. 
\begin{enumerate}
\item Imprecise default data: The default value at the initial point in MIPS.
\item Precise simulation data: The exact solution collected in the numerical solver, i.e., MIPS. We take the ground-truth value as the precise data.
\end{enumerate}
Our sensitivity study first checks the convergence criteria in the MIPS code and collects those variables critical to the simulation converge, namely $X, \lambda$, $\mu$, and $Z$.
We use these (im-)precise data as initial points to test the importance of each variable in two aspects, i.e., the impact on success rate and speedup.
Here, \textit{success rate} refers to the ratio of those initial solution can reach the convergence criteria to the total number of input problems. \textit{Speedup} is time acceleration, namely the rate of actual solving time to the exact solving time in MIPS.

Then, we include eight test systems\footnote{ Refer to Table~\ref{tab:model_conf} for a more detailed illustration of test systems.} and generate $10,000$ samples for each system by varying input loads to analysis the impact of using different initial points. 
Initializing the four variables with precise and imprecise types, we have $2^4$ combinations to analyze the contribution of each variable.
Table~\ref{tab:Ablation} shows the results \textcolor{revision}{of} 16 combinations. For each combination, we use ``0'' and ``1'' to indicate the imprecise data and precise data respectively. To calculate the success rate and speedup, we take a baseline, the combination where all variables using imprecise default data. 


\textcolor{revision}{Table \ref{tab:Ablation} reveals that precision improvement on these initial variables does not always bring benefits to simulation performance in terms of success rate and speedup. A high improvement on precision might even decrease the success rate of the simulation. For example, the baseline case \uppercase\expandafter{\romannumeral1} (using all default parameters) has a success rate of 100\%, while improving the precision on the feature $Z$ (e.g., the case \uppercase\expandafter{\romannumeral2}) alone leads to failed convergence at most input problems. 
Hence, blindly building a NN model to numerically approach those precise values may reduce success rate and lose simulation performance.}

We further analyze the performance results in Table \ref{tab:Ablation} and summarize some interesting observations.
\begin{itemize}
    \item \textbf{Observation 1}: Using precise $X$ leads to a $100\%$ success rate (see case \uppercase\expandafter{\romannumeral9}), while the features $X$, $\lambda$, $\mu$ and $Z$ jointly contribute to high speedup see case \uppercase\expandafter{\romannumeral 16}). 
    \item \textbf{Observation 2}: The contribution of  $Z$ to the success rate and speedup strongly depends on whether $\mu$ use precise data. For example, the success rate is dropped down when involve a precise $Z$ without a precise $\mu$ (see case \uppercase\expandafter{\romannumeral12} with respect to \uppercase\expandafter{\romannumeral10}).
    
    \item \textbf{Observation 3}: The contribution of $\lambda$ to the success rate and speedup is independent of whether the other features use precise data or not. For example, the success rate of initialing with a precise $\lambda$ \textcolor{revision}{does} not change with/without a precise $X$ or $\mu$ and $Z$ (see case \uppercase\expandafter{\romannumeral5}, \uppercase\expandafter{\romannumeral8}, \uppercase\expandafter{\romannumeral13} and \uppercase\expandafter{\romannumeral16}).     
    
    \item \textbf{Observation 4}: Features $X$, $\lambda$, $\mu$ and $Z$ have implicit dependency.   Improving accuracy of one feature cannot guarantee the overall performance improvement to the success rate and speedup. For example, improving the accuracy of $\lambda$, $\mu$, or $Z$ on a precise $X$ can not guarantee the improvement of success rate and speedup (comparing  case \uppercase\expandafter{\romannumeral9} with cases \uppercase\expandafter{\romannumeral10}, \uppercase\expandafter{\romannumeral11} and \uppercase\expandafter{\romannumeral13}.)
\end{itemize}
Observations 1 and 4 indicates that \textcolor{revision}{the analyzed features} are highly inter-dependent, which can be targeted on multi-task prediction.
We \textcolor{revision}{build} a MTL model to enable the information sharing for the inter-dependency.
We decide feature priority by \textit{dependency between features}, namely, the contribution of a feature to success rate and speedup is changed with a variation of another feature. For example, since $\lambda$, $\mu$ and $Z$ have dependency on $X$, we should make $X$ very accurate. We give $X$ the highest priority. 

Observation 2 and 3 reveals features show differences on dependency. Some features (e.g., $\lambda$) are relatively independent while others (e.g., $\mu$ and $Z$) have dependency, which implies customized design for different feature prediction should be considered in modeling. Also, we observe that $\lambda$ is an equality factor while $\mu$ and $Z$ contribute to inequality together in Eqn~\ref{eqn:largran}. Such information from physical law validates feature dependency and maybe can be used to deal with the dependency in model training inversely.
 
Driven by these observations, we introduce an interactive learning model for multi-objective modeling (discussed in Section~\ref{sec:multitask}) and impose domain knowledge to strength physical understanding for prediction quality (discussed in Section~\ref{sec:physics_info}).

\section{An interactive learning model}
\label{sec:multitask}

In this section, we develop a MTL model to enable multitask prediction. Based on the observation from sensitivity study, we implement domain specific design through prioritizing features and enforcing a physics-dependent hierarchy in the MTL model. After that, we depict the details of MTL parameters. 
\vspace{-3pt}
\subsection{Multitask Learning}
Multitask learning is an inductive transfer learning method~\cite{caruana1997multitask, torrey2010transfer}. A MTL model is typically composed of shared layers and task-specific layers. 
Unlike using multiple separate models for each task, the MTL model enables us to share information from common layers while customizing specific layers for corresponding tasks. The training signals of different tasks can be learned as inductive biases to facilitate the learning of all tasks, which achieves a unification of \textcolor{revision}{the} shared information and \textcolor{revision}{the} task-specific information.

\textbf{Information-sharing in shallow layers. }
Observation 1 reveals that there is correlation among these four features and we would like to leverage these relations in our model.
To incorporate this correlation, we utilize information sharing in MTL by parameter sharing and loss sharing.

\begin{itemize}
    \item \textbf{Parameter sharing.} The common layers share the same weights and topology between tasks. By sharing parameters, tasks share low-level semantic information to complement domain knowledge with each other. 
    Meanwhile, parameter sharing can alleviate the risks of overfitting due to the noise brought by multiple tasks.
    
    \item \textbf{Loss sharing.} \textcolor{revision}{The} tasks to predict $X, \lambda, \mu, Z$ share a common loss function to update training gradient in the MTL. The minimal loss of different tasks are usually in different positions. By sharing losses, the MTL passes information and avoids being trapped in a local optimal. 
\end{itemize}

\vspace{-3pt}
\textbf{Task-specific learning in deeper layers. }
Besides \textcolor{revision}{the} features $X, \lambda, \mu, Z$ \textcolor{revision}{being} correlated, observations 2 and 3 shows \textcolor{revision}{that} specific design for different feature prediction should be considered. 
In task-specific layers, we introduce specific model topologies (estimators) for each task based on \textcolor{revision}{the} task demands. For example, a task requires a positive output. We apply a rectified linear unit (ReLU), a type of activation functions, at the last layer to bound the output always positive.

Figure~\ref{fig:MTL} shows the topology of the proposed MTL.
Given a power network topology, we use the power load (including both the active part $P_d$ and reactive part $Q_d$) as \textcolor{revision}{model inputs} and estimate seven tasks (four variables in $X$). In the MTL, the shared layers are extracting information from different tasks, while the task-specific layers (estimators) utilize customized designs to generate their own dedicated results.

\subsection{Domain-Specific Design}

\begin{figure}[!t]
  \centering
  \includegraphics[width=0.9\linewidth]{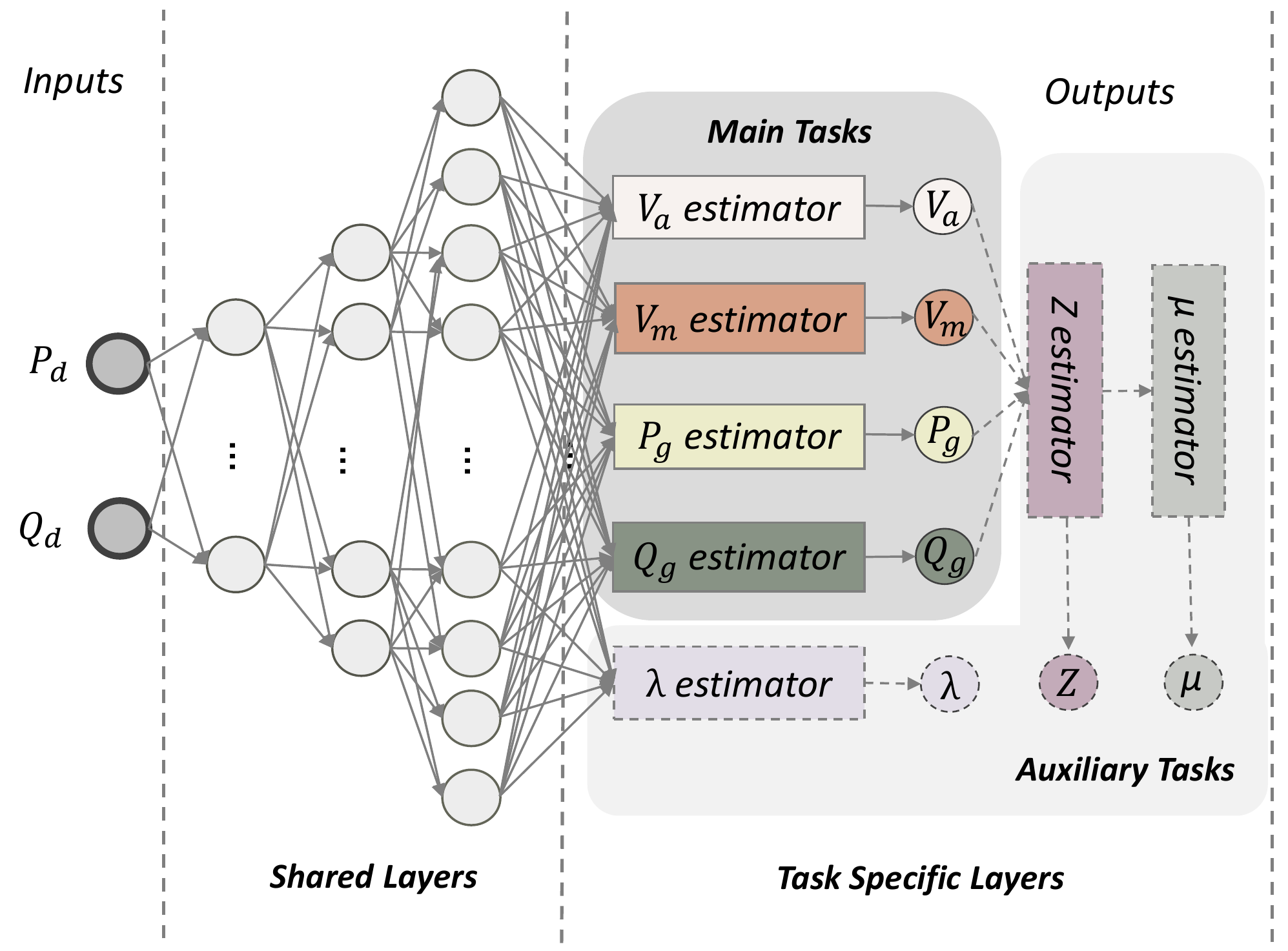}
    \vspace{-10pt}
  \caption{Topology of the MTL.}
  \vspace{-18pt}
  \label{fig:MTL} 
\end{figure}

Besides using shared layers and task-specific layers, we introduce a domain-specific design into MTL: this design is driven by our observations on (1) the contribution difference of the four features \textcolor{revision}{to} success rate and speedup and (2) \textcolor{revision}{the} dependency between features. The domain-specific design includes two techniques, feature prioritization and a physics-dependent hierarchy, discussed as follows.

\textbf{Feature prioritization.} Observation 1 shows that precise $X$ guarantees the success of simulation convergence, while precise $\lambda$, $\mu$, and $Z$ contribute to simulation acceleration. We prioritize the four features by specifying the prediction of $X$ as the main task while the prediction of other three features as auxiliary tasks. The auxiliary tasks are used as an augmentation to provide additional information for the main task. 
Through ``eavesdropping'' the main tasks, the auxiliary tasks interact with the main task implicitly. More importantly, learning the direct solution $X$ in the main task gives the user high simulation quality while estimating the Lagrangian factors ($\lambda$, $\mu$, and $Z$) in the auxiliary tasks maximize performance speedup. 
Technically, we apply ``detach()" operation~\cite{paszke2019pytorch} for these auxiliary tasks periodically. \textcolor{revision}{The} detach operation blocks the gradient back-propagation to the shared layers (which are contribute to the main task $X$). In other word, we set a knob of detach operation to alternately train the main task or the entire model.
In particular, the MTL focuses on improving main tasks when we activate the detach operation, while facilitates the interaction between main tasks and auxiliary tasks when the detach operation is disabled.

\textbf{A physics-dependent hierarchy. }
Observations 2 and 3 reveal the dependence between $Z$ and $\mu$ and the independence of $\lambda$, respectively.
We find these observations are consistent with the computation order in the solving process. In particular, the simulation process takes the order of (1) computing $X$ and $\lambda$; (2) computing $Z$ based on $X$; and (3) computing $\mu$ based on $X$ and $Z$. This is consistent with the existing work~\cite{sereeter2019comparison,sereeter2019linear}.

To fully exploit the benefit of information sharing, we enforce a physics-dependent hierarchy in MTL. As shown in Figure~\ref{fig:MTL}, we first infer the main task $X$($V_a$, $V_m$, $P_g$, $Q_g$) and an independent auxiliary task $\lambda$. Then, we predict task $Z$ based on $X$. After that, we estimate $\mu$ based on the predicted $Z$.  

\subsection{Details on Multitask Learning Model}


\textcolor{revision}{In this section,} we present details about \textcolor{revision}{the} topology parameters, \textcolor{revision}{the}  loss function and \textcolor{revision}{the} pre-processing method of the MTL model. We use a power grid system of 300 buses as an example, but the MTL modeling method is general for any other power grid systems. Figure~\ref{fig:MTL} generally depicts the model topology.

The shared layers take the power load $P_d$ and $Q_d$ at each bus as input, totaling 600 inputs. There are five fully-connected layers as the shared layers. We set the numbers of neurons in the five layers as 600, 720, 840, 960, and 1080 respectively. 
The five fully-connected layers extract shared features and feed them to seven specific estimators (four in $X$ and $\lambda$, $\mu$, $Z$), each of which is a fully-connected network customized for a task.
We use ReLU as the activation function to increase the model nonlinearity.
We use a variant of $L_1$ loss~\cite{janocha2017loss}, the Charbonnier loss, as our loss function. This is a supervised loss function calculating the difference between each of the predicted output variables $v$ and the corresponding ground-truth value $v_{gt}$ collected in the MIPS solver. Our loss function is defined as follows. 
\begin{equation}
\label{eqn:charb_loss}
    L = \frac{1}{|\mathcal{V}|}\sum_{v\in \mathcal{V}}  W_v \sqrt{(v - v_{gt})^2 + \epsilon ^2}
\end{equation}
where $\mathcal{V}$ is a set consisting of $V_a$, $V_m$, $P_g$, $Q_g$, $Z$, $\lambda$ and $\mu$; $W_v $ is a weight for a task $v$ and $\epsilon$ is a small constant for numerical stability. We set $\epsilon$ as $1e-9$ in our study.

\section{Physics-Informed Learning}
\label{sec:physics_info}

The solution in power grid simulation must respect several physical constraints, such as power generation, line flow, and bus voltage constraints. Incorporating these constraints into the MTL model not only improves model accuracy but also increases the model interpretability.

In general, the constraints are classified into hard and soft constraints. 
The \textit{hard constraint} includes some strict bounds on the variable ranges in applications; 
The \textit{soft constraint} includes domain knowledge to improve model accuracy, such as physical principles, conservation laws, and others gained from theoretical or computational studies. 
We introduce four objective functions to incorporate domain knowledge and impose those constraints by minimizing the objective functions.

\subsection{Embedding AC Nodal Power Balance Equations} 
The power grid simulation includes a power flow equality constraint shown in Eqn.~\ref{eqn:3} to make the simulation of the power grid system stable and make the simulated solution feasible. We integrate the AC nodal power balance equations (Eqn.~\ref{eqn:3}) into the objective function $f_{AC}$ to guide model training. 
\begin{equation}
\label{eqn:fAC}
\begin{split}
    f_{AC}=  \vert P_d + P_{bus}(Y_{bus}, V_a, V_m) - P_i(C_g,P_g)\vert + \\
    \vert Q_d + Q_{bus}(Y_{bus}, V_a, V_m) - Q_i(C_g,Q_g)\vert 
 \end{split}
\end{equation}
The above objective function bridges model inputs ($P_d,Q_d$), outputs ($X$, $\lambda$, $\mu$, $Z$) and the physics information ($C_g, Y_{bus}$) of power networks to yield quantitatively better physical connection. In particular, the generator connection matrix $C_g$ and the bus admittance matrix $Y_{bus}$ are critical information determined by the physical network of power system.
Eqn.~\ref{eqn:3} shows the AC power system keeps stable only if the power generation equals to the power load. 
In the objective function $f_{AC}$, we calculate the differences between power load and generation and minimize the difference approaching to zero.

\begin{figure}[!t]
  \centering
  \includegraphics[width=1.0\linewidth]{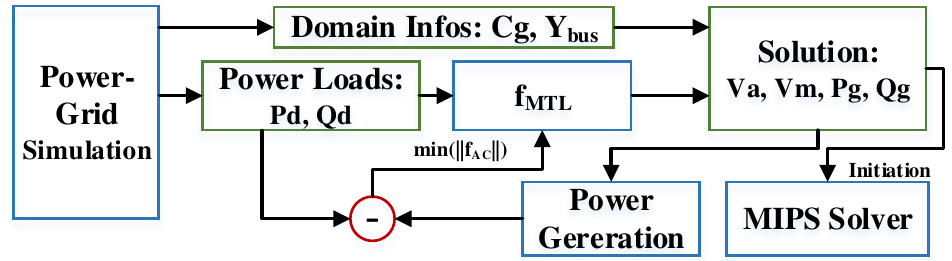}\
    \vspace{-20pt}
  \caption{Embedding AC physical laws in MTL training }
    \vspace{-15pt}
  \label{fig:ACtraining} 
\end{figure}

Figure~\ref{fig:ACtraining} shows how the objective function $f_{AC}$ works in MTL training. In the power grid simulation, we utilize domain information $C_g$ and $Y_{bus}$, which provide power-grid bus topology and resistance information respectively.  Power loads ($P_d$ and $Q_d$) are the input fed to the MTL to produce solutions $V_a, V_m, P_g, Q_g$. We integrate the AC power balance law (Eqn.~\ref{eqn:fAC}) to calibrate the training \textcolor{revision}{loss} in $f_{MTL}$. In particular, we calculate the power generation based on the prediction solution $X$ and domain information $C_g$ and $Y_{bus}$, and subtract power generation from the power loads. We then calculate the difference between the power loads and power generation, and try to minimize the difference within the objective function $f_{AC}$. The only block with training parameters is the $f_{MTL}$ and all blocks are differentiable.

The above training process is driven by the predicted data and facilitates the prediction inversely. Such a data-driven architecture $f_{AC}$ mitigates the risk of obtaining infeasible solutions, such as those predicted solution misled by the noise of training data and violating the basic AC power balance law. 
Embedding the objective function $f_{AC}$ has two significant benefits. First, since the information of model inputs is limited to predict its outputs, we incorporate non-trivial data-augmentation as a complementary to increase prediction accuracy. Second, we can efficiently perform transfer learning with fewer training data even if the typology of power network is modified, e.g., a transmission line in the power-grid bus is suddenly broken. With this, we can improve our MTL prediction and facilitate the solution robustness.

\subsection{Guarding Inequality Constraints} 

The AC-OPF formulation includes two inequality constraints: one is explicit, quantitatively bounding $X$  by $X_{min}<X<X_{max}$ (Eqn.~\ref{eq:1d}), and the other is implicit, limiting branch flow by $H(X)>0$ (Eqn.\ref{eq:1c}). For the implicit inequality, we impose physics information $C_g$ and $Y_{bus}$ to calculate the branch flow state $H(X)$ and check if the $H(X)$ violates the bounds. 
We utilize exponential functions to punish the overflow error in these inequality constraints and force the prediction to be bounded by the expected, normalized range. Eqn.~\ref{eq:ineq_constraint} shows how we use the exponential functions.
In the equation, we build an objective function $f_{ieq}$ to incorporate the inequality equations as a penalty loss.
\begin{equation}
\label{eq:ineq_constraint}
    f_{ieq} = e^{-H(X)} + e^{(X-X_{max})} + e^{(X_{min}-X)}
\end{equation}
Where $X$ is a predicted feature in MTL. Once the predicted $X$ violates the inequality constraints, for example, $H(X_k)<0$, the overflow error will be visibly shown up in the objective function $f_{ieq}$ and calibrated through  backpropagation in the training phase. Hence, the main task $X$ is restricted in a quantitative way to improve simulation quality.




Guarding inequality constraints in our model mitigates the overflow error in inequality constraints while facilitates the feasibility of model prediction. 

\subsection{Optimization of Cost Function} 
The ultimate goal of the AC-OPF is to minimize the cost function $f(X)$ \textcolor{revision}{(Eqn.~\ref{eq:1a})}. We explore the physics information in $f(X)$ to construct an objective function $f_{f(X)}$ and minimize the loss between the predicted cost and ground truth cost.
\begin{equation}
  f_{f(X)} = |f(X) - f_0|
\end{equation}
where $f_0$ is the ground-truth value of the cost collected by the numerical solver, MIPS. 
Feature $X$ is our model prediction. In $f(X)$, we utilize the characteristics of energy consumption on generators to calculate the predicted cost $f(X)$.
Then, the objective function $f_{f(X)}$ calibrates the predicted cost $f(X)$ with the ground-truth cost $f_0$ to reach the optimal solution. 

\subsection{Implying Lagrangian Conservation} 
In Eqn.~\ref{eqn:largran}, the AC-OPF problem can be solved as the equality constraints $G(x)=0$ and slacked inequality constraints $H(x)+Z=0$ approach zero.
Here, we apply two ways to imply the Lagrangian formulation into MTL training. 
First, we reconstruct the inequality and equality constraints as soft constraints, which is imposed in the loss function to guide the training. 
Then, we refer the variable bounds $Z>0$ and $\mu>0$ as hard constraints and apply an activation function to strictly bound model prediction.
We construct an objective function $f_{Lag}$ to guide the training subject to the soft constraints.
\begin{equation}
    f_{Lag}= |\mathbf{\lambda}^\intercal G(X)|+|\mathbf{\mu}^\intercal(H(X)+Z)|
\end{equation}

We incorporate the hard constraints during the training phrase by projecting predictions onto a region induced by the constraints. In particular, we first pre-process the raw data of ground truth into the normalized range $[0,1]$. Then, we apply a ``sigmoid'' activation function at the last layer to bound the output range of $Z$ and $\mu$ to be positive and into the same range $[0,1]$. The above techniques provide hard upper and lower bounds on prediction and guarantee its feasibility.

Incorporating $f_{AC}$ and $f_{ieq}$ improves the feasibility and robustness of the simulation solution $X$;  Incorporating $f_f{(X)}$ improves the accuracy of $X$;  Incorporating $f_{Lag}$ can optimize auxiliary tasks $\lambda, \mu $ and $Z$.  Hence, we arithmetically compose these objective functions into the loss function $L$(Eqn.~\ref{eqn:charb_loss}).
\begin{equation}
   L_{total} = L+ L_{eqn} + L_{ieq} +  L_{lag} + L_{f(X)} 
\end{equation}
The $L_{total}$ efficiently combines supervised learning (with ground-truth labels) and unsupervised learning (without ground-truth labels) to guide the MTL training. By doing so, we maintain the prediction accuracy while increase the model feasibility and interpretability. 

 \section{Evaluation}
\label{sec:eva}

We evaluate our framework by examining its impacts on performance and simulation quality of power grid simulation.

\noindent\textbf{Platform.} 
\textcolor{revision}{We conduct all experiments on an NVIDIA DGX-1 cluster with 16 nodes, and each node is equipped with two Intel Xeon E5-2698 v4 CPUs (40 cores running at 2.20GHz) and 8 NVIDIA TESLA V100 (Volta) GPUs.}
We use CUDA 10.1/cuDNN 7.0~\cite{chetlur2014cudnn} to run NNs on NVIDIA GPUs. We use Pytorch for model training and inference.

\noindent\textbf{Matpower.}
Matpower 6.0 is an open-source Matlab
power system simulation package~\cite{zimmerman2016matpower}, which is used widely in research and education for AC- and DC- power flow simulations. 
The default OPF solver, i.e., Matlab Interior Point Solver~(MIPS), is a high-performance primal-dual interior-point solver.

\noindent\textbf{Load Sampling.}
We sample the loads within $[(1 - t) \times P_{di},(1 + t) \times P_{di}]$ uniformly at random, where $P_{di}$ is the default power load at the $i$-th bus, and $t$ is the variation percentage, i.e., 10\% in this paper, consistent with state-of-the-art~\cite{pan2019deepopf, guha2019machine, zamzam2019learning}.

\noindent\textbf{Input Datasets.}
\textcolor{revision}{To comprehensively evaluate the performance, we use five power networks in Table~\ref{tab:model_conf} as test systems.} 
We generate 10,000 input problems for each test system, in which 8,000 of them are for training and 2,000 for validation. 
These samples are fed into Matpower to produce the optimal solutions as the supervision ground truth signal.
Uniform sampling is applied to avoid over-fitting issues common in generic DNN approaches~\cite{gittens2016revisiting}. 
\begin{table}[!t]
\centering
\caption{\textcolor{revision}{Configurations in IEEE bus systems.}} 

\label{tab:model_conf}
\begin{tabular}{l p{0.9cm}p{0.9cm}p{0.9cm}p{1.1cm}p{1.1cm}}
\arrayrulecolor{revision}\hline
\textcolor{revision}{Problem size}  & \textcolor{revision}{14-bus} & \textcolor{revision}{30-bus} & \textcolor{revision}{57-bus} & \textcolor{revision}{118-bus} & \textcolor{revision}{300-bus} \\
\arrayrulecolor{revision}\hline
\textcolor{revision}{Buses}  &\textcolor{revision}{14} & \textcolor{revision}{30} & \textcolor{revision}{57} & \textcolor{revision}{118} & \textcolor{revision}{300} \\
\textcolor{revision}{Generators}& \textcolor{revision}{5} & \textcolor{revision}{6}  & \textcolor{revision}{7} & \textcolor{revision}{54} & \textcolor{revision}{69} \\
\textcolor{revision}{Branches} & \textcolor{revision}{20} & \textcolor{revision}{41} & \textcolor{revision}{80} & \textcolor{revision}{185} & \textcolor{revision}{411} \\
\textcolor{revision}{\#$\lambda$} & \textcolor{revision}{29} & \textcolor{revision}{61} & \textcolor{revision}{115} & \textcolor{revision}{237} & \textcolor{revision}{601}\\
\textcolor{revision}{\#$\mu (Z)$} & \textcolor{revision}{48} & \textcolor{revision}{166} & \textcolor{revision}{142} & \textcolor{revision}{452} & \textcolor{revision}{876}\\ 
\hline
\end{tabular}
\end{table}


\subsection{\name Performance Evaluation}
\label{sec:eva_overall}

In our approach, we use \name to generate a high-quality initial condition for the MIPS solver, thereby drastically reducing the overall time-to-solution. 
We introduce the following performance metric to calculate the achieved speedups by \name:
\begin{equation}
  \label{eqn:speedup}
  SU = \frac{T_{MIPS}}{T_{MTL}+T'_{MIPS} +T_{MIPS}\times(1-SR)}  
\end{equation} 
where $T_{MIPS}$ represents the solving time when using the traditional approach with MIPS, $T_{MTL}$ represents the inference time of the MTL model, and $T'_{MIPS}$ represents the convergence time in MIPS initializing with the output of \name. \textcolor{revision}{$T_{MIPS}\times(1-SR)$ calculates the restart execution time with the default initial point in MIPS if the simulation fails.}
SR represents the overall success rate, $SR = N_{suc}/N_{total}$, where $N_{suc}$ represents the number of problems successfully solved by MTL and $N_{total}$ represents the total number of input problems.
\textcolor{revision}{Whenever the initial condition provided by \name does not lead to the simulation converge successfully, we fall back to the traditional MIPS solver to guarantee the final convergence.
Hence our method always provides 100\% guarantee on simulation convergence,} though it might come at an additional cost of re-executing overhead in the workflow.
\begin{figure}[!t]
  \centering
  \includegraphics[width=1.0\linewidth]{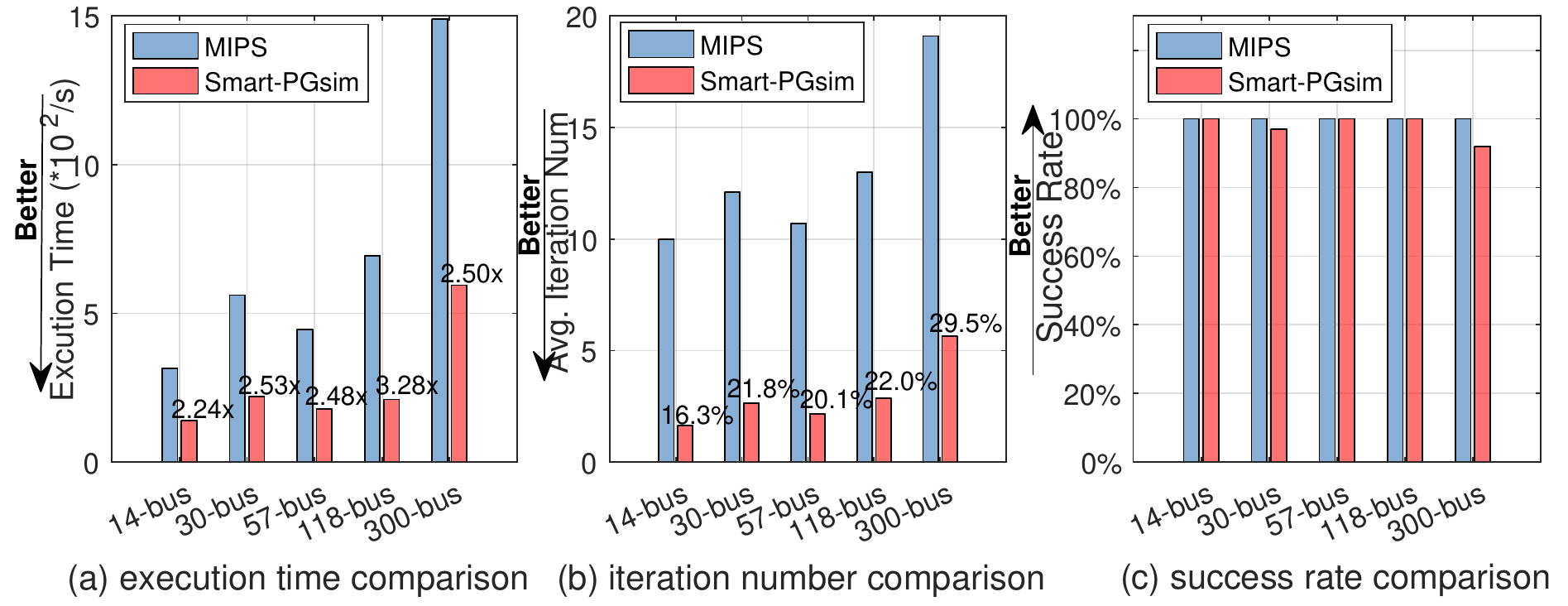}
 \caption{Comparison of three aspects between MIPS and \name.}
  \label{fig:over_speedup} 
\end{figure}
\begin{figure}[!t]
  \centering
  \includegraphics[width=1.0\linewidth]{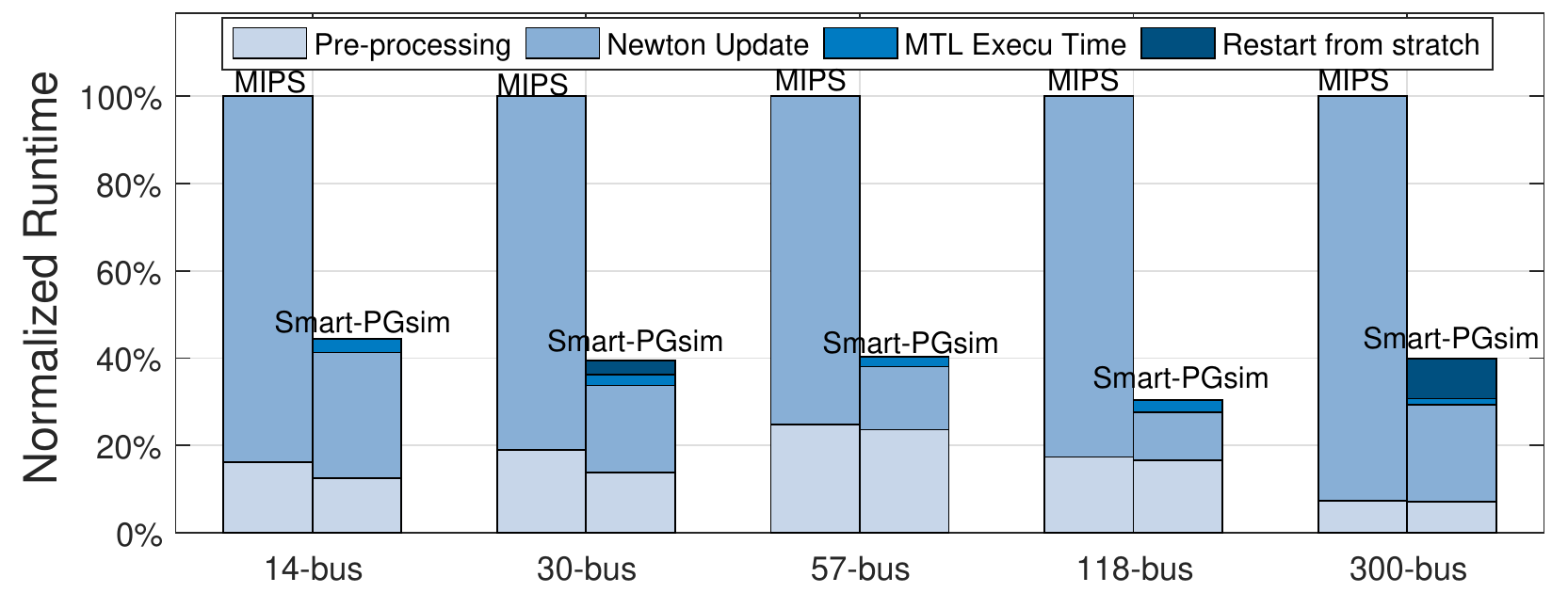}
  \caption{Execution time breakdown}
  \label{fig:per_bd} 
\end{figure}

Figure~\ref{fig:over_speedup}(a) compares the execution time of the traditional numerical simulation performed with MIPS and that of our framework in terms of the $SU$ metric described above.
Each test system is run on $2,000$ input problems. Performance measurements of \name comprise the end-to-end runtime, including the time to produce the warm-start points in MTL, the convergence time in MIPS with the warm-start points, and the restart execution time in MIPS if the simulation fails.
In the Figure~\ref{fig:over_speedup}(a), we also label the speedup at the top of \name bar. The \name speedups over MIPS observed in the plot are considerable, ranging from nearly a $2.24\times$ speedup up to over a $3.28\times$ speedup.
Furthermore, the performance benefit of \name is more evident as the size of power networks increases, which indicates a notable potential in accelerating large-scale power grid systems.
{It is important to note that using \name as warm-start for MIPS generates the same solution as produced by MIPS directly.}
Figure~\ref{fig:over_speedup}(b) presents the average iteration number of MIPS and \name across different test systems, in which we only consider the iteration acceleration produced by \name.
We measured the average iteration number during the convergence process until the terminate criteria are reached.
The iterative process is the most computationally intensive  part of the power grid simulation.
We also label the ratio of the \name iteration number to the MIPS iteration number on \name bar.
The results in Figure~\ref{fig:over_speedup}(b) show that \name dramatically reduces the number of iterations required to converge, taking only $16.3\%$ to $29.5\%$ iterations of previous conduction (MIPS). 
The accelerated convergence drives the overall performance improvement of \name.

\subsection{Performance Breakdown}
To further explore the performance improvement provided by \name, Figure~\ref{fig:per_bd} shows the runtime breakdown of MIPS and \name, normalized to the overall runtime where the problems run with MIPS.
The pre-processing refers the execution time of problem construction and data preparing for power grid simulation, in which MIPS and \name show almost the same processing time.
The Newton update represent the execution time spent in Newton iteration. 
\name have extra overheads about the inference time of the MTL model for generating warm-start solution and the restart time with failure cases.
\textcolor{revision}{Note that we restart the failed cases with the default setting in the numerical solver MIPS to guarantee the final convergence.}
As Figure~\ref{fig:per_bd} depicts, \name is effective at reducing the time spent on the convergence, i.e., Newton Update. 
\name demonstrates significant performance improvement for the tested input problems despite the extra overhead introduced by the MTL model.

\begin{figure}[!t]
  \centering
  \includegraphics[width=1.0\linewidth]{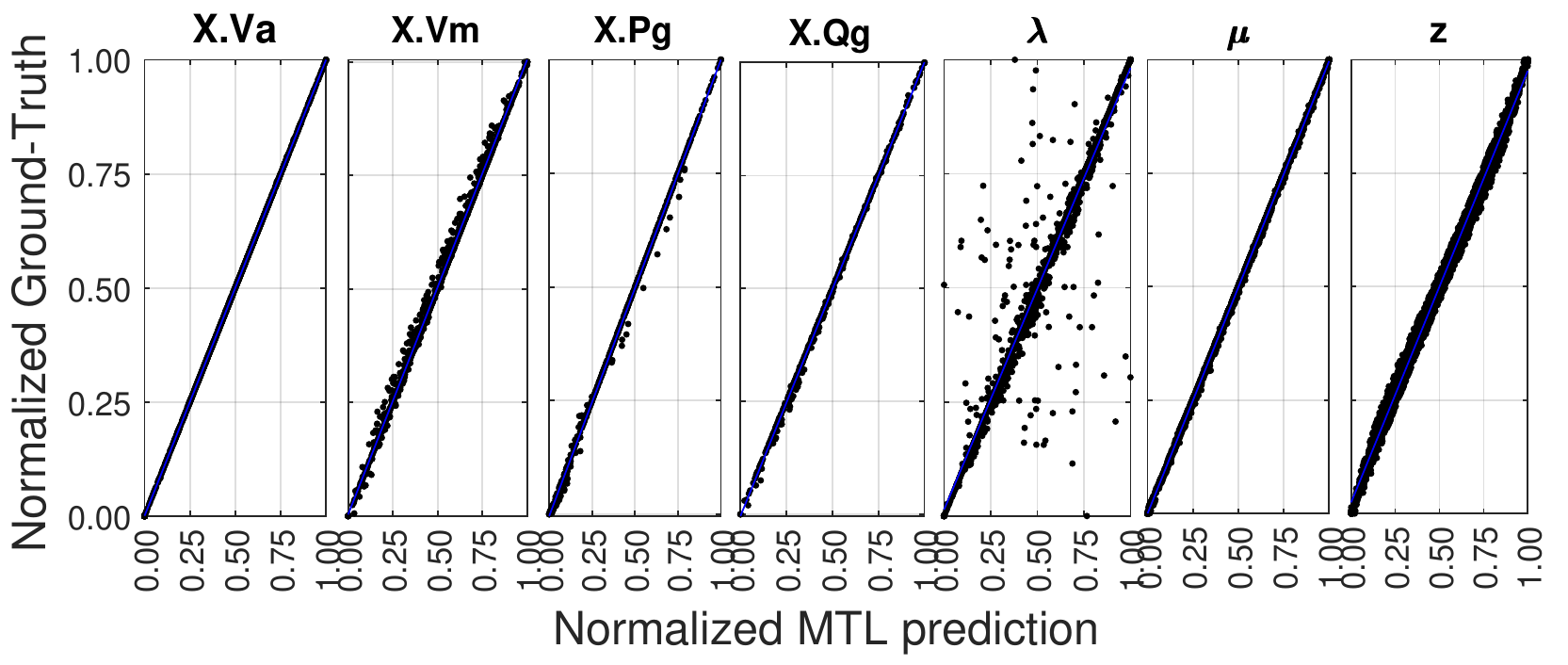}
  \caption{Prediction accuracy of each feature used in the proposed MTL model.}
  \label{fig:err_ana} 
\end{figure}

\subsection{Prediction Accuracy}
\label{sec:eva_acc}
Figure~\ref{fig:over_speedup}(c) compares MIPS and \name  in terms of the success rate, in the case in which we do not restart \name after a failed execution. \textcolor{revision}{As we discuss above, \name guarantees \textit{100\% success rate in practice} by re-executing those computations that do not provide high-enough accuracy, while still considerably outperforming MIPS execution.} The success rate is how many input problems can converge successfully in the simulation. Figure~\ref{fig:over_speedup}(c) reveals that \name leads to a high percentage of success rate in all case. \name provides 100\% success rate on 14-bus, 57-bus, 118-bus while maintains a relatively high success rate as 97\% and 92\% on 30-bus, 300-bus respectively. 

Figure~\ref{fig:err_ana} presents the prediction accuracy of each feature used in \name.
We compare the accuracy of warm-start points predicted by \name with the exact solution in MIPS (Ground-truth). 
The prediction and ground-truth are normalized to the range [0,1].
The x-axis is the predicted value of \name and the y-axis is the ground-truth value. If the prediction of \name is perfect, all points should be lie on the $y=x$ line.
There is negligible accuracy lost in the prediction of $X.Va$, $X.Vm$, $X.Pg$, $X.Qg$, $\mu$ and $z$.
For $\lambda${}, there is a larger variation in the predicted values representing over-prediction and under-prediction.
Such variation in $\lambda$ is acceptable because $\lambda$ is the equality constraints factor in Eqn.~\ref{eqn:largran}, which will not affect the final convergence if the equality constraints are satisfied.

\subsection{Efficiency of Multitask Learning and Physical Constraints}

In this section, we analyze the effectiveness of multitask learning and imposing physics constraints. 
First, we develop a model of multiple separate NNs without information sharing.  
For peer comparison, we use the same number of layers and neurons as MTL model in the multiple separated networks.
Then, to show the efficiency of physical constraints, we remove physics constraints in MTL model as a comparison. 

\begin{figure}[!t]
  \centering
  \includegraphics[width=1.0\linewidth]{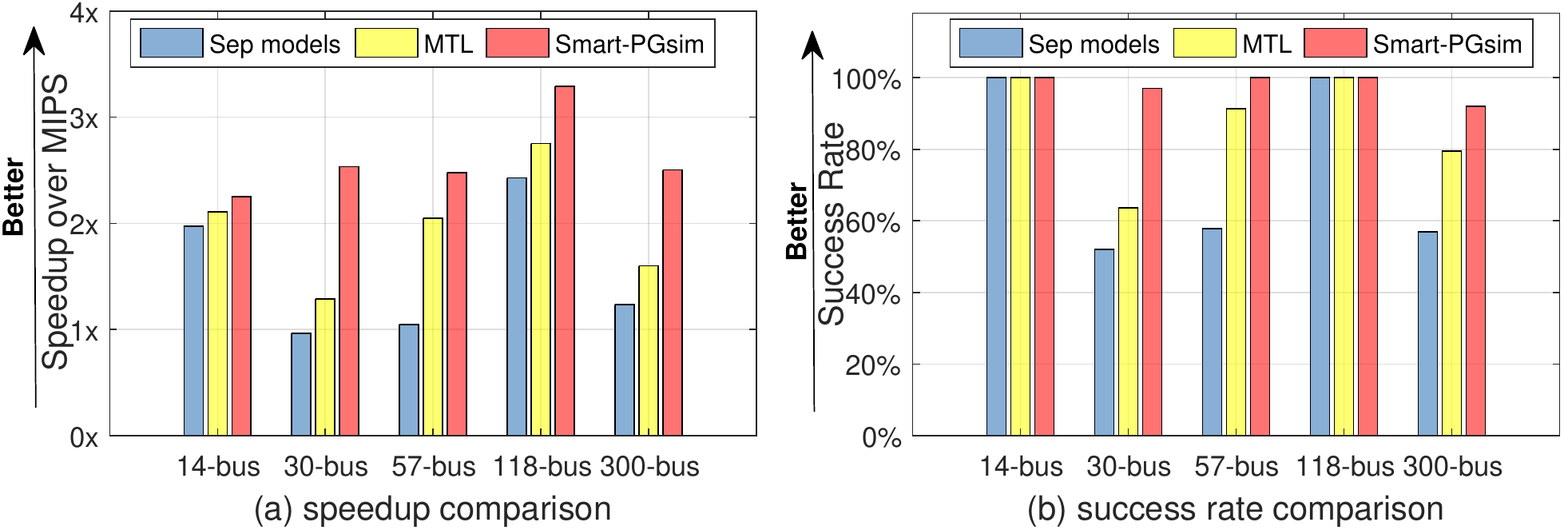}
  \caption{Performance comparison}
  \label{fig:per_comp} 
\end{figure}

\begin{figure}[!t]
  \centering
  \includegraphics[width=0.95\linewidth]{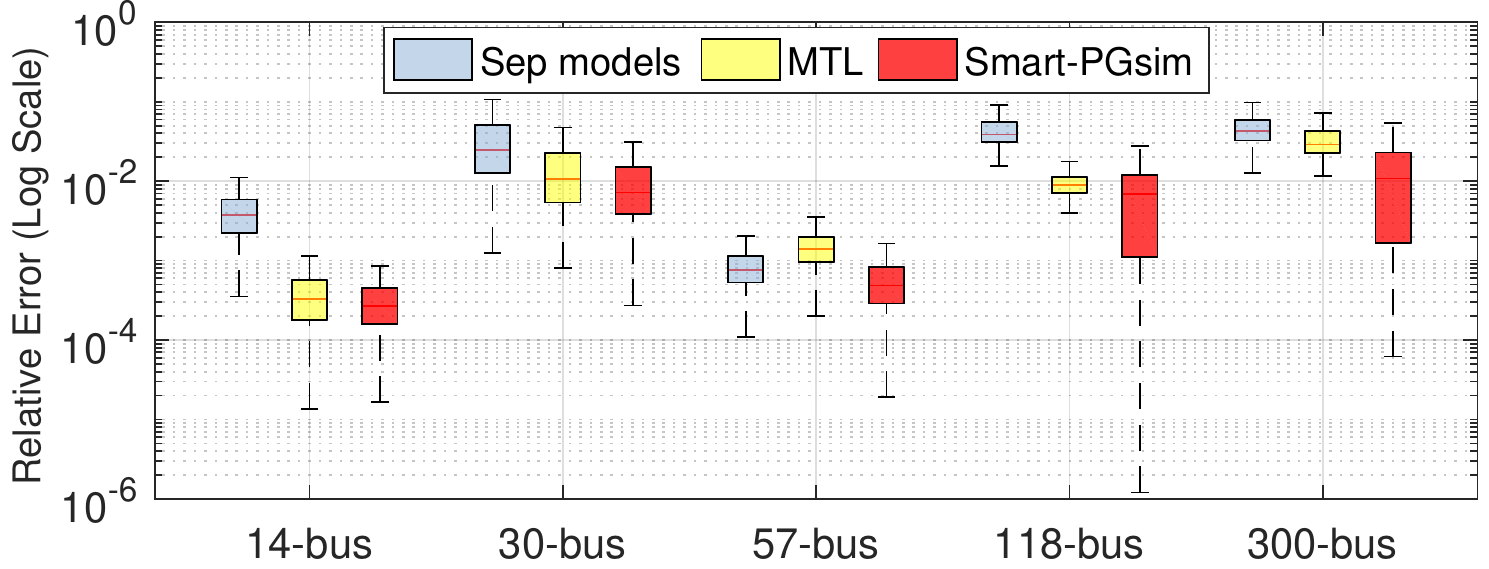}
  \caption{Accuracy comparison.}
  \label{fig:phy_constraints} 
\end{figure}

Figure~\ref{fig:per_comp}(a) shows the speedup comparison with the multiple separated NNs, MTL model and \name. 
Here, ``MTL'' refers to the multitask learning model without physical constraints whereas ``\name'' refers the multitask learning model with physical constraints.
Note that all speedup are measured with our performance metric $SU$ in Eqn.~\ref{eqn:speedup}.  
Figure~\ref{fig:per_comp} shows that the performance of the speedup $SU$ and the success rate $SR$ are significantly improved by the multitask learning and incorporation of the physical constraints.  
MTL provides notable speedup and success rate improvement over the multiple separated models, average speedup of $1.36\times$ and $22.5\%$ success rate improvement.
In particular, the multiple separated NNs show inefficiency on the test system 30-bus with a $0.96\times$ speedup, in which the 52.0\% success rate produce a soaring overhead on restart. 
Adding the physical constraints further improve the speedup and success rate by $40\%$ and $18.3\%$ over MTL. 
In summary, our proposed framework \name, a multitask model with physical constraints offers the highest average speedup and solution feasibility.

Moreover, Figure~\ref{fig:phy_constraints} presents box-plots\footnote{In the box-plots, the boxes are bounded by 25-th and 75-th percentiles of the variables; The central marks of the boxes indicate the median~\cite{dawson2011significant}. } to show the results of the prediction accuracy in different models.
We use relative error $RE=|V_{predict} - V_{gt}|/{V_{gt}}$ to measure the prediction accuracy. 
$V_{predict}$ refers the prediction values of MTL and $V_{gt}$ refers the exact solution in MIPS (i.e., ground-truth). The lower relative error means the prediction is more accurate and closer to the groud-truth.
We draw two observations from Figure~\ref{fig:phy_constraints}: (1) The prediction provided by \name has the lowest average error with all five test systems; (2) Most of the predictions in \name is under the error line of $10^{-2}$, which shows \name consistently produces prediction within $1\%$ relative error.  
These two observations reveal that \name can provide more \textit{consistent acceleration} than multiple separate models and MTL, which is crucial for dealing with widely diversified input problems in real-time.

\subsection{\textcolor{revision}{Scalability Analysis on Multi-Node Systems}}
\label{sec:eval_scale}

\begin{figure}[!t]
  \centering
  \includegraphics[width=1.0\linewidth]{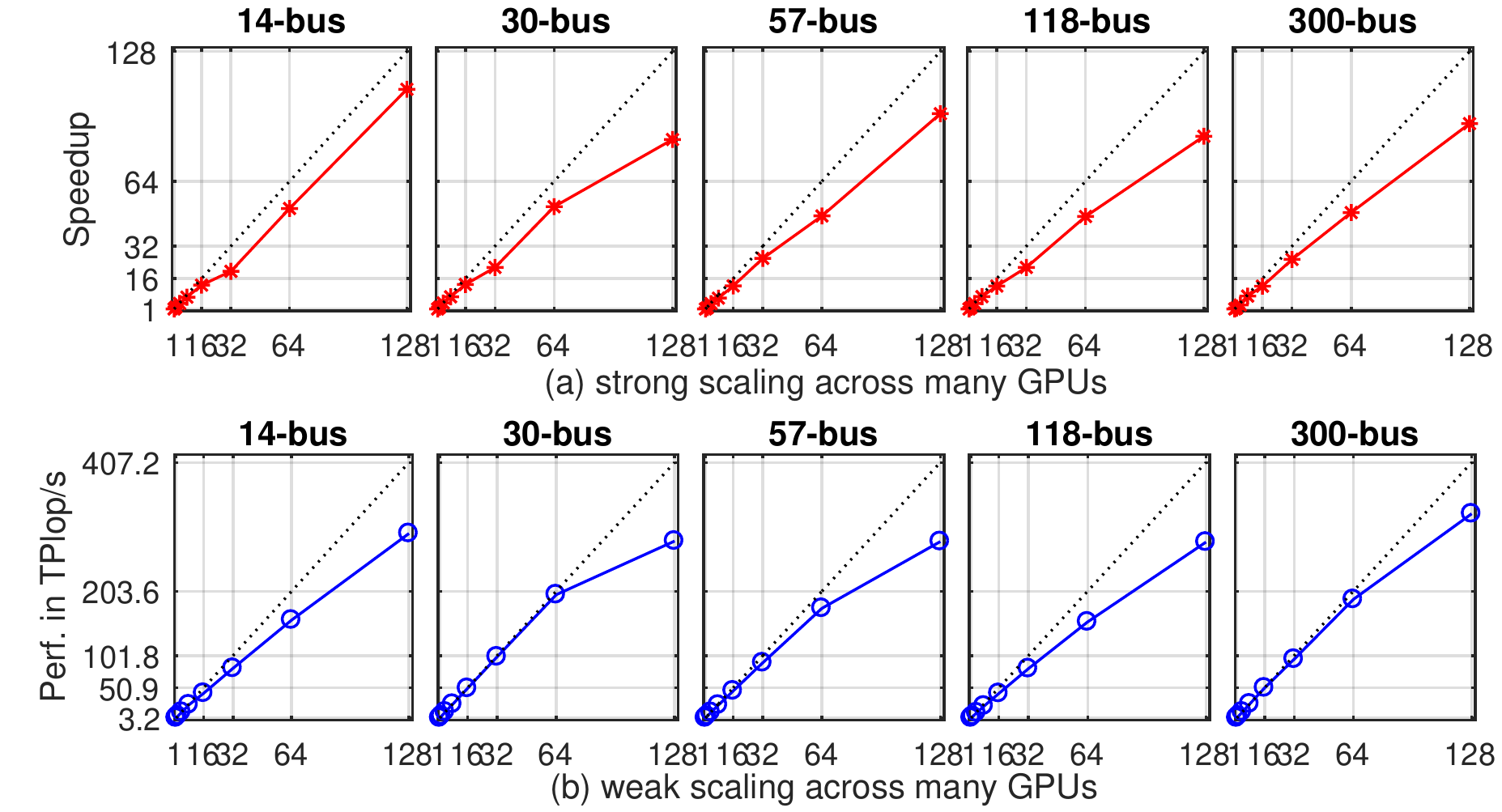}
  \caption{\textcolor{revision}{Scaling across many GPUs}}
  \label{fig:mul_sca} 
\end{figure} 
\textcolor{revision}{As we stated earlier, the AC-OPF problem is solved many times per day by power operators throughout the life of the power grid. Additionally, the real-life problem is further complicated by the uncertainty involved by equipment security, the reliability of alternative power sources (solar, eolic, and hydro power), and the stability of power generators and power transmission lines. Considering all those factors together is generally referred to solving Security-Constrained ACOPF (SC-ACOPF)~\cite{chiang2015solving,8442590} and it originates very large and complex uncertain scenario trees that need to be analyzed to maintain the robustness of the global solution, i.e, an optimal solution that survives under all uncertain scenarios. From a computational perspective, these scenarios are largely independent (although some similarities can be exploited to reduce computational requirements) and result in a computational problem that is largely embarrassingly parallel and, thus, inherently scalable on parallel computers (e.g., assigning a batch of scenarios to each compute node, and then assigning a set of scenarios from the batch to each GPU).}


\begin{table}
\normalsize
 \caption{Prediction Performance Comparison.}
\label{tab:model_perf}
\begin{adjustbox}{width=1.0\linewidth}
\begin{tabular}{c cccccc}
\toprule 
\multirow{3}{*}[-0.33em]{\textbf{Zamzam's~\cite{zamzam2019learning}}}            & \textbf{Test system} & --              & \textbf{39-bus} & \textbf{57-bus} & \textbf{118-bus} & \textbf{--} \\
\cmidrule(lr){2-7}
& \textbf{SF }& -- & $15.38\times$ &$9.49\times$ & $7.97\times$& --\\
&\textbf{$L_{cost}$}& -- & $0.326\%$ & $0.457\%$& $0.821\%$& -- \\
\midrule
\multirow{3}{*}[-0.33em]{\textbf{\name}} & \textbf{Test system} & \textbf{14-bus} & \textbf{30-bus} & \textbf{57-bus} & \textbf{118-bus} & \textbf{300-bus} \\
\cmidrule(lr){2-7}
&\textbf{SF} & $21.17\times$  & $40.19\times$ &$21.72\times$ & $36.15\times$& $105.64\times$ \\
&\textbf{$L_{cost}$}& $0.007\%$ & $0.074\%$ & $0.040\%$& $0.003\%$& $0.008\%$ \\
\bottomrule
\end{tabular}
\end{adjustbox}
\end{table}

\textcolor{revision}{While the focus of this work is mainly on accelerating each AC-OPF instance of a larger SC-ACOPF problem by providing high-quality initial conditions for the numerical solver, one can easily imagine that speedup similar to the ones reported in Section~\ref{sec:eva_overall} can be expected for the SC-ACOPF problem. To verify such assertion, we conducted experiments on a 16-node compute cluster, where each node is an NVIDIA DGX-1 equipped with eight NVIDIA V100 GPUs (128 GPUs in total). We study both \textit{strong scalability} and \textit{weak scalability}. Strong scaling is measured with a fixed number of scenarios, while weak scaling linearly increases the number of scenarios with respect to the number of processors. 
\name is expected to generate an initial solution for each scenario.
In these experiments, we use data parallelism for scaling out the \name workload, in which each GPU has an identical copy of the entire network and each computes results for a separate set (the local batch) of input scenarios.}
\textcolor{revision}{We focus on scaling Smart-PGSim, which emulates the use case where there are needs to generate initial solutions for a large number of scenarios for the SC-ACOPF problem.}


\begin{figure*}[!t]
  \centering
  \includegraphics[width=\linewidth]{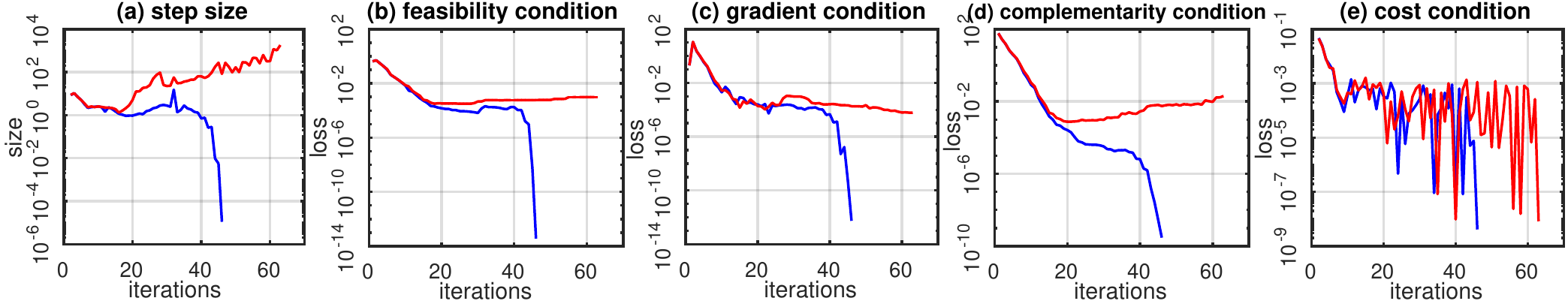}
  \caption{\textcolor{revision}{ The asymptotic convergence of the tracking loss along the iterations} }
  \label{fig:iter_trj} 
\end{figure*} 

\textcolor{revision}{
Figure~\ref{fig:mul_sca}(a) shows strong scaling behavior for each of the five test systems with a fixed problem size (10k scenarios) from 1 to 128 GPUs.  The black dotted lines in the plots represent ideal scaling for data parallelism. As expected, increasing the number of GPUs naturally leads to a higher speedup and shows an almost linear tendency. However, the speedup is not linear.}

\textcolor{revision}{
Such a non-linear speedup is caused by our work distribution strategy: While our distribution algorithm has been designed to equally distribute scenarios between GPUs,  communication effects can skew this balance. Specifically, when running in a node with 8 GPUs, we first copy the MTL model and data to the first GPU device and then copy it to the other GPUs leveraging GPUDirect and NVLINK, which generates some load imbalance that translates into efficiency loss.}

\textcolor{revision}{Figure~\ref{fig:mul_sca}(b) shows similar results for the weak scaling experiments, where the number of scenarios increases from 10k to 1,280k when increasing the number of GPUs from 1 to 128 (10k scenarios per GPU). The scalability shown in the plots for weak scaling is better,  compared to that for strong scaling. This is because the weak scaling uses larger problems which amortizes the load imbalance problem, but we still notice similar issues as the strong scaling experiments.}

\textcolor{revision}{
Overall, \name scales up to 128 GPUs: for the test system of ``300-bus'' (the largest system we evaluated), we achieve a peak performance of 604.7 TFLOPS and a sustained performance of 326.1 TFLOPS, reaching 43\% of the peak performance of Volta V100 (double precision).}

\subsection{Comparison with Prior Work}

Previous work~\cite{zamzam2019learning,guha2019machine} use ML to directly replace the exact solver to achieve a high speedup.
For a fair comparison with the state-of-the-art method, i.e., Zamzam et al.'s model~\cite{zamzam2019learning} that leverages DNN for prediction, we assume that the prediction of \name is the final solution, effectively replacing the entire solving computation.
In Table~\ref{tab:model_perf}, we compare  performance and optimality loss to what has been used in Zamzam's model.
Cost deviation measures simulation quality. We donate a speedup factor (SF) to measure the computational improvements: $SF = \frac{1}{n}\sum_{i=1}^{n} {(T^{MTL}_{i}}/{T^{MIPS}_{i})}, $ where $T^{MTL}_i$ refers the execution time of the MTL and $T^{MIPS}_i$ represents the execution time of the numerical solver (MIPS) for each input problem $i$.
We measure the loss using the average fractional difference between the predicted cost $C'$ and the true cost $C$:
\(L_{cost} = \frac{100\%}{n} \sum_{i=1}^{n}\left|1-C'_{i}/ C_{i}\right|\).
The results presented in Table~\ref{tab:model_perf} show that our framework outperforms the state-of-the-art even in the case in which we directly use \name output as final solution of the computation. \name achieves an average $44.97\times$ speedup which provides $310.7\%$ improvement comparing the average speedup of Zamzam's model ($10.95\times$). Moreover, \name decreases the cost loss $12.16\times$ by average comparing with Zamzam's model.
Although  \name show significant improvement over the state-of-the-art, we remark that the solutions produced by both models might not satisfy the strict requirements of power-grid simulations, hence our approach further refines \name output in the traditional solver MIPS at the back end and achieving high-quality solutions, although with reducing the speedup.

\section{Discussions}

\textcolor{revision}{In this section, we discuss the generality of our techniques and analyze solving processes with and without convergence.}

\subsection{Generality of Proposed Approach} 
\textcolor{revision}{The three major techniques, including sensitivity study (Section~\ref{sec:sensi_study}), multi-tasking learning (Section~\ref{sec:multitask}), and incorporating domain knowledge (Section~\ref{sec:physics_info}), can be broadly applied to many scientific HPC applications, and are not limited to the optimization problem in power grid simulations. In this section, we highlight some potential application of our techniques to other scientific applications.}

\textcolor{revision}{\noindent\textbf{Fluid dynamic simulation} aims to study the flow of fluid materials. Smart-fluidnet model~\cite{dong2019adaptive} is a convolutional NN model to accelerate fluid simulation. We can build a multitask learning model by predicting the output velocity filed $\vec{u}$ as a main task and pressure field $p$ as an auxiliary task since $p$ impacts the fluid movement ($\vec{u}$). Moreover, we can incorporate the incompressibility condition, $\bigtriangledown \cdot \vec{u} = 0$, as a physical constraint regularized as a soft constraint $L_{loss}=\bigtriangledown \cdot \vec{u}$.}

\textcolor{revision}{
\noindent\textbf{Molecular Dynamics (MD) simulation.} The DPMD model~\cite{lu202086} is an NN model to accelerate MD simulation.  This model can be enhanced by using the techniques presented in this work by developing a multi-task model where the main task predicts the potential energy and an auxiliary task predicts the symmetry-preserving descriptor. Also, the potential energy should be positive, which can be enforced as a hard constraint.}

\textcolor{revision}{
\noindent\textbf{Cosmology modeling.} CosmoFlow~\cite{mathuriya2018cosmoflow} is an NN model to predict three cosmological parameters that can be directly implemented as multi-task learning. The Cosmic Microwave Background~\cite{ade2016planck} can be enforced as a hard constraint to bound the projection range of modeling.}

\subsection{Analysis of Diverging Cases}

\textcolor{revision}{The solving process for the AC-OPF problem can fail to converge. Figure~\ref{fig:iter_trj} shows the inconvergence process given a bad initial solution and compares it with the convergence process given a good initial solution. Figure~\ref{fig:iter_trj} shows the variance of step size and four convergence conditions across iterations. The step size $|\Delta x|$ refers to the length of the updating step during the simulation; The four conditions are used to determine if the simulation is converged in each iteration.}

\textcolor{revision}{Figure~\ref{fig:iter_trj} shows that, for the case with bad initial solution, the step size rapidly increases. Accordingly, the four convergence conditions remain relatively stable without being able to converge. For the case with good initial solution, the step size and three conditions (feasibility, gradient and complementary) decrease quickly. We notice that the cost condition goes through great variance in both cases, which makes it  difficult to correlate to convergence.} 

\textcolor{revision}{The step size is critical to determine the direction to explore to find the optimal solution. If the initial solution is bad, the solving process aims to use a larger step size to find a promising direction. However, using a large step size could lead a failure of convergence (Figure~\ref{fig:iter_trj}.a).}

\textcolor{revision}{As our results show, it is difficult to guess whether the numerical solver will converge based on the first iterations: both good and bad initial conditions behave similarly during the initial iterations of the power grid simulation and there is no clear indication that some computation will later fail. Given this complexity, we resort to re-initialize and re-execute the numerical solver from the beginning without employing the initial conditions generated by the MTL model.
Overall, as our results demonstrate, even considering restart time, \name still significantly outperforms state-of-the-art solutions.}

\section{Conclusions}
\label{sec:conclusion}

Using NN to approximate and/or accelerate high performance computing applications has shown promising results. However, how to effectively apply a NN to those applications is still an open question. The approximations introduced by the NN models need to be carefully analyzed, so that the simulation quality in the application is not lost and even improved; at the same time, the execution time of the application should be reduced after applying NN. 
In this paper, we apply a NN to accelerate a specific power grid simulation problem, AC-OPF. As a simulation to solve complex nonlinear optimization problems based on iterative numerical methods, AC-OPF raises challenges on simulation robustness (i.e., ensuring the optimality of the simulation solution for various input problems) and respecting the physical constraints imposed by the power flow. We introduce a framework, \name, that facilitates the construction of a NN model by studying the impact of the output accuracy on simulation convergence and execution time and automatically imposing the physical constraints.  Using a novel multitask-learning NN model generated by \name, we produce high-quality initial solutions for 10,000 input problems. Based on those solutions, the AC-OPF simulation reduces simulation time by an average of $2.60\times$ (up to $3.28\times$) without losing the optimality of the solution.

\section{Acknowledgement}

This research is supported by the U.S. Department of Energy (DOE) Advanced Scientific Computing Research (ASCR), U.S. National Science Foundation (CNS-1617967, CCF-1553645 and CCF-1718194), award 74756, Co-design of Reconfigurable Accelerators for Sparse, Irregular Computations Underlying Machine Learning and Graph Analysis, award 66150,  CENATE - Center for Advanced Architecture Evaluation, Chameleon cloud and XSEDE resource.


\normalem
\bibliographystyle{unsrt}
\bibliography{main}

\end{document}